\documentclass[12pt]{iopart}
\usepackage{epsfig}
 
\usepackage{bm}% bold math
\usepackage{iopams}

\begin{document}
\title[Inhomogeneities in HTSc - perspective of two-component model]{Real
 space inhomogeneities in high temperature 
superconductors: the perspective of two-component model 
}

\author{J.Krzyszczak$^1$, T.Doma\'nski$^1$, K.I.Wysoki\'nski$^{1,2}$, 
R.Micnas$^3$ and S.Robaszkiewicz$^3$} 
\address{$^1$ Institute of Physics and Nanotechnology Center,\\ 
       M.\ Curie - Sk\l odowska University, 20-031 Lublin, Poland}
\address{$^2$ Max Planck Institut f\"ur 
             Physik komplexer Systeme,  
             D-01187 Dresden, Germany}
\address{$^3$ Faculty of Physics, A. Mickiewicz University, 
Umultowska 85, 61-614 Pozna\'n, Poland}

\ead{karol@tytan.umcs.lublin.pl} 

\date{ \today}

\begin{abstract}
The two-component model 
of high temperature superconductors in its real space version
 has been solved using  Bogoliubov-de Gennes 
equations. The  disorder in the electron and boson subsystem 
has been taken into account. It strongly modifies the
superconducting properties and leads to local
variations of the gap parameter and density of states.
The assumption that the impurities mainly modify 
boson energies offers natural explanation
of the puzzling positive correlation between the positions of 
impurities and  the values of the order parameter found in the scanning
tunnelling microscopy experiments.  
\end{abstract}

%71.10.Fd 	Lattice fermion models (Hubbard model, etc.)
%74.20.-z 	Theories and models of superconducting state
%74.25.Op 	Mixed states, critical fields, and surface sheaths
%74.40.+k 	Fluctuations (noise, chaos, nonequilibrium 
%superconductivity, localization, etc.)
%74.81.-g 	Inhomogeneous superconductors and superconducting systems
\maketitle

\section{Introduction}
Recent years have witnessed the interesting series
of experiments on high temperature superconductors (HTS).
These include {\it inter alia} scanning tunnelling spectroscopy measurements 
which revealed atomic scale inhomogeneities on the surface of BiSrCaCuO
\cite{cren00,pan01,howald2001,lang02,mcelroy05} and 
the quantum  oscillations 
\cite{Doiron-Leyraud07,bangura08,chakravarty2008} which
are signatures of the well developed Fermi surface. 
All these discoveries clearly indicate that after more 
than 20 years of research,  high temperature superconductors (HTS) still
present a challenge and are the objects of intensive 
experimental \cite{damascelli2003,devereaux2007,sonier2008,basov2005}
and theoretical studies \cite{norman2005,lee2006,ogata2008,lee2008}.
  
The scanning tunnelling microscopy (STM) experiments are mainly 
performed on Bi family of  HTS \cite{fisher2007}.
These materials cleave easily and it is straightforward to obtain
atomically clean surfaces predominantly with  BiO surface layer.
This makes  Bi superconductors also interesting materials
to study by means of angle-resolved photoemission spectroscopy (ARPES)
\cite{damascelli2003}. 
The strength of STM technique consists in its high energetic and
spatial  resolution. With STM one is able to study  changes 
of local properties within the atomic distances.    
The poor screening, short superconducting coherence 
length of cuprate superconductors and 
low-dimensionality of the electronic structure  contribute 
to large spatial fluctuations 
in normal  and superconducting state properties.

STM  measures the quasiparticle excitations, existing at
a given energy. From   
the  differential conductance spectra  one deduces {\it e.g.}  
 local values of superconducting gap. The study 
of HTS reveals a number of unexpected features in the STM spectra.
Among  those ``universally" observed there was 
nearly a homogeneous  structure of conductance spectra 
at low energies despite 
strong variations at higher energies and in particular,
 inferred gap values \cite{howald2001}, the asymmetry
({\it i.e.} $G(-V) \ne G(V)$) of the differential 
conductance \cite{renner1996} $G(V)=dI/dV$   and characteristic 
dip-hump structure \cite{matsuda2003}.

Already in the early days of HTS,  the measured STM spectra 
have shown high degree of disorder \cite{liu1991,chang1992}.
The subsequent more precise experiments allowed 
for the analysis of the results not only in real, 
but also in reciprocal space.
It is important to note that with the Fourier transformed STM data, 
the possibility appeared to reconstruct the energy 
gap $\Delta({\bf k})$, which turned
out to be in agreement \cite{davis2003} with the ARPES measurements.  
The study of the spatial maps resulted in discovery 
that the values of the superconducting gaps
are {\it positively} correlated with positions 
of oxygen dopants thus supporting the earlier claims 
that electronic inhomogeneities and atomic 
disorder are interrelated. 
Since then inhomogeneities are thus commonly attributed
to  the  presence of disorder  \cite{balatsky2006}.

The development \cite{Gomes07} of the so called ``lattice-tracking spectroscopy"
has enabled investigation of various correlations between 
the local system properties even at significantly 
different temperatures. 
 This resulted in finding that not only the 
 positions of dopant oxygen atoms correlate  \cite{mcelroy05, mashima2006}
 with local values of the superconducting gap, but 
that also the latter property is positively correlated 
with local density of states (LDOS) measured in the 
normal state at a temperature well above that of the
superconducting transition. This correlation indicates that
normal state contains information about the interactions responsible
for superconducting instability of the system \cite{pasupathy2008}.

The purpose of this paper is to  study theoretically  
 local properties  of HTS relevant to the  STM experiments. 
We describe the superconductor by  the two-component 
boson-fermion (BF) model \cite{ranninger1985,robaszkiewicz1987}. 
For the homogeneous system the Hamiltonian can be written
as 
\begin{eqnarray}
{\hat{\cal H}}^{{BF}} &=&
\sum_{\vec k,\sigma} (\varepsilon(\vec k)-\mu) \hat{c}_{\vec k\sigma}^{\dagger} \hat{c}_{\vec k\sigma} +
 \sum_{\vec q} \left( E_{\vec q}^{B}   
- 2\mu \right) \hat{b}_{\vec q}^{\dagger} \hat{b}_{\vec q}   \nonumber
\\
&+& \frac{1}{\sqrt{N}} \sum_{\vec k, \vec q} g_{\vec k,\vec q} \left( \hat{b}_{\vec q}^{\dagger} 
\hat{c}_{\vec q-\vec k \downarrow} \hat{c}_{\vec k \uparrow} + 
\hat{c}_{\vec k\uparrow}^{\dagger} \hat{c}_{\vec q- \vec k\downarrow}^{\dagger} \hat{b}_{\vec q} \right).
\label{BF_hamilt0}
\end{eqnarray}
The $c_{\vec k}(c_{\vec k}^{\dagger})$ operators refer to single 
fermions and annihilation (creation) operators of bosons
are denoted by $b_{\vec q}(b_{\vec q}^{\dagger})$. 
The d-wave character  of the order parameter in superconducting state  
is taken into account by assuming for two dimensional lattice 
$g_{\vec k}= g[\cos(k_xa)-\cos(k_ya)]$.

Coexisting local pairs (bosons) and electrons are described
by the Hamiltonian (\ref{BF_hamilt0}) which has been proposed
on phenomenological basis \cite{ranninger1985} but can also be derived 
from generalized periodic Anderson model \cite{robaszkiewicz1987} or as
a low energy limit of cluster states \cite{altman2002}.
Its relevance to HTS and other unconventional superconductors 
has been demonstrated in numerous works of various groups 
\cite{micnas1990,friedberg1990,ranninger1995,enz1996,geshkenbein1997,kostyrko1997,castroneto2001,romano2001,noce2002,micnas2002,domanski2004,micnas2004,micnas2007}. 
In a broader context   this
phenomenological two-component pairing scenario  plays a role in a great variety
of different physical situations. (i) It applies to  resonant pairing in
cold atomic fermion gases via a Feshbach resonance \cite{bloch2008}, 
(ii) it captures the 
salient features of Anderson's resonating 
valence bond scenario \cite{altman2002} 
and (iii) it permits to draw conclusions about
the pairing mechanism, if given by local 
dynamical lattice fluctuations,
as was the original motivation for  proposing this model \cite{ranninger1985}.

As in this work we are  concerned with the  description of 
inhomogeneities and correlations observed experimentally, we generalize
the  Hamiltonian (\ref{BF_hamilt0}) to the real space 
 and make the parameters random numbers. It means that 
disorder  is responsible for various kinds of inhomogeneities. 
In the absence of {\it ab-intio} calculations \cite{altman2002} 
of how the boson and fermion parameters behave in an impure system \cite{petit2009}  
 we study a few scenarios of their possible changes.

It should be stressed that we are not averaging over disorder as experiments
measure local values for a given (fixed) configuration. 
It turns out  that with the BF model a  number 
of experimental   correlations found in STM  can
be understood with  reasonable assumptions about the role of disorder.
Here we shall be mainly interested in the maps of local density of states,
the shapes of differential conductance curves as the functions of energy
(applied bias) as measured at  different sites and the values of energy gaps.
Preliminary studies of this model have  been  
presented recently \cite{krzyszczak2008}.
In literature there exist numerous attempts 
to  describe theoretically various aspects 
of the STM spectra in HTS cuprates including {\it inter alia} 
 inhomogeneities 
\cite{martin2001,wang2001,z-wang2002,nunner2005,andersen2006,maska2007},
asymmetry of the differential conductance \cite{domanski2003} and 
dip-hump structures \cite{gabovich2007}. 

The organization of the rest of the paper 
is as follows. In section 2 we introduce the two-component model, 
discuss the approach and compare the 
calculations for a small homogeneous system with bulk one.
The results for the model with disorder in the electron and boson 
subsystems are presented in section 3 whereas main conclusions
in section 4.

\section{The model and approach}
We start with the  two-component
model in a real space where fermions interact {\it via}
short range forces with the hard-core charged and localized bosons.
This boson-fermion model \cite{robaszkiewicz1987,micnas2004} 
is represented by the Hamiltonian  
\begin{eqnarray}
{\hat{\cal H}}^{{BF}} &=& \nonumber
\sum_{i,j,\sigma} t_{i j} \hat{c}_{i\sigma}^{\dagger} \hat{c}_{j\sigma} +
 \sum_{i\sigma} \left( V_{i}^{{imp}} -
 \mu \right) \hat{c}_{i\sigma}^{\dagger} \hat{c}_{i\sigma} + \sum_{i} \left( E_i^{B}   
- 2\mu \right) \hat{b}_{i}^{\dagger} \hat{b}_{i}\\
&+& \sum_{i,j} \frac{g_{i j}}{2} \left[ \hat{b}_{i}^{\dagger} 
(\hat{c}_{i\downarrow} \hat{c}_{j\uparrow} -\hat{c}_{i\uparrow} \hat{c}_{j\downarrow})+ \hat{b}_{i} 
(\hat{c}_{j\uparrow}^{\dagger} \hat{c}_{i\downarrow}^{\dagger}-
\hat{c}_{j\downarrow}^{\dagger} \hat{c}_{i\uparrow}^{\dagger}) \right],
\label{BF_hamilt}
\end{eqnarray}
where   $i$ and $j$ denote the lattice sites, 
$\hat{c}_{i,\sigma}^{\dagger}$ ($\hat{c}_{i,\sigma}$) 
stand for creation (annihilation) operator  
of fermion at the site $i$ with spin $\sigma$. 
 $\hat{b}_{i}^{\dagger}$ and $\hat{b}_{i}$ are the creation
 and annihilation operators of hard-core bosons at the site $i$.
$\mu$ stands for the chemical potential of the system and $t_{i j}$ 
are hopping integrals. $g_{ij}$ is the electron-boson scattering
(charge exchange). $V^{imp}_i$ is the local value of the 
fermionic energy level (scattering
potential due to impurities) and $E^B_i$ denotes the position of the
bosonic level.
  We shall discuss the systems on a square
lattice with hopping parameters between the nearest neighbors to be
denoted  $t$, which we use 
as the energy unit.

\subsection{Bogoliubov - de Gennes equations}
 
To account for the short coherence length and
d-wave symmetry of the superconducting order
parameter we assume that the boson-fermion coupling $g_{i j}$ 
takes on non-zero value  for the nearest neighbor 
sites $<i,j>$ only and equals +g if $j=i\pm\vec{x}$ 
and -g if $j=i\pm\vec{y}$.

Application of the standard Hartree-Fock-Bogoliubov decoupling of the type
$ \hat{b}_{i}^{\dagger} \hat{c}_{i\downarrow} \hat{c}_{j\uparrow}
\approx \langle \hat{b}_{i}^{\dagger}\rangle \hat{c}_{i\downarrow} 
\hat{c}_{j\uparrow}
+\hat{b}_{i}^{\dagger}\langle\hat{c}_{i\downarrow} \hat{c}_{j\uparrow}\rangle$
leads to $\hat{H}^{{BF}}=
\hat{H}^{{B}}+\hat{H}^{{F}}$, where 
$\hat{H}^{{B}}$ is the single site bosonic Hamiltonian 
\begin{equation}
\hat{H}^{{B}}= \sum_{i} \left( E_i^{B}   
- 2\mu \right) \hat{b}_{i}^{\dagger} \hat{b}_{i}
+ \sum_{i}  \left( \hat{b}_{i}^{\dagger}  \chi_i + 
\hat{b}_{i} \chi_{i}^*\right) 
\end{equation}
with the parameter $\chi_{i} = \displaystyle \sum_{<j>}  (g_{i j}/2)
 <\hat{c}_{i,\downarrow} \hat{c}_{j,\uparrow}-\hat{c}_{i,\uparrow} \hat{c}_{j,\downarrow}>$
depending on the fermionic degrees of freedom.  
$\hat{H}^{{F}}$ is the Hamiltonian describing 
the disordered fermionic subsystem \cite{domanski2002}. 
\begin{eqnarray}
\hat{H}^{{F}}& =& 
\sum_{i,j,\sigma} t_{i j} \hat{c}_{i\sigma}^{\dagger} \hat{c}_{j\sigma} +
 \sum_{i\sigma} \left( V_{i}^{{imp}} -
 \mu \right) \hat{c}_{i\sigma}^{\dagger} \hat{c}_{i\sigma} \\ \nonumber
&+& \sum_{i,j}  \left[ \Delta_{ij}^* 
(\hat{c}_{i\downarrow} \hat{c}_{j\uparrow}-\hat{c}_{i\uparrow} \hat{c}_{j\downarrow}) + \Delta_{ij} 
(\hat{c}_{j\uparrow}^{\dagger} \hat{c}_{i\downarrow}^{\dagger}-
\hat{c}_{j\downarrow}^{\dagger} \hat{c}_{i\uparrow}^{\dagger}) \right].
\end{eqnarray}
The  standard statistical approach allows for an exact 
solution of the boson Hamiltonian $\hat{H}^{{B}}$.  
For a given realization of disorder one finds the following
equations for the local number of bosons   $<\hat{b}_{i}^{\dagger} \hat{b}_{i} >$
and the local boson average $<\hat{b}_{i}>$
\begin{eqnarray}
<\hat{b}_{i}^{\dagger} \hat{b}_{i} > &=& \frac{1}{2} - \frac{E_i^{B} 
 - 2\mu}{4\epsilon_{i}} \tanh \left( \frac{\epsilon_{i}}
 {k_{B}T} \right), \label{n_b} \\
<\hat{b}_{i}> &=& - \frac{\chi_{i}}{2\epsilon_{i}} 
\tanh \left( \frac{\epsilon_{i}}{k_{B}T} \right).
\label{ord_pars}
\end{eqnarray}
We have denoted $\epsilon_{i} = \sqrt{ \left( \frac{E^{B}_{i} 
- 2\mu}{2} \right)^{2} + |\chi_{i}|^{2}}$  and $\Delta_{ij} 
= g_{i j} <\hat{b}_{i}>$/2. It is $\Delta_{ij}$ 
which couples two subsystems as follows from 
equation (\ref{eq_BdG}) below.

The fermion part has the standard BCS-structure 
and we diagonalize it by the Bogoliubov-Valatin transformation,
which introduces the new quasiparticle 
operators \cite{ketterson1999}
$\hat\gamma_{l,\downarrow},\hat\gamma_{l,\uparrow}$ and their adjoint
counterparts $\hat\gamma^{\dagger}_{l,\downarrow},
\hat\gamma^{\dagger}_{l,\uparrow}$
\begin{eqnarray}
\hat c_{i,\downarrow}=\sum_l \left[u_i^l\hat\gamma_{l,\downarrow} + 
v^{l*}_i\hat\gamma^{\dagger}_{l,\uparrow}\right], \\ \nonumber
\hat c_{i,\uparrow}=\sum_l \left[u_i^l\hat\gamma_{l,\uparrow} - 
v^{l*}_i\hat\gamma^{\dagger}_{l,\downarrow}\right]. 
\label{bog-val}
\end{eqnarray}

This yields the following Bogoliubov-de Gennes 
equations \cite{krzyszczak2008,domanski2002}  
\begin{eqnarray}
\sum_{j}
\left(
\begin{array}{c}
t_{i j}+\left( V_{i}^{{imp}}-\mu \right)\delta_{ij} ;  \tilde\Delta_{ij} \\
\tilde\Delta_{ij}^{*} ; 
 -t_{i j}-\left( V_{i}^{{imp}} -\mu \right) \delta_{ij}
\end{array}
\right)\left(
\begin{array}{r}
u^{l}_{j} \\
v^{l}_{j}
\end{array}
\right)=
E^{l}\left(
\begin{array}{c}
u^{l}_{i} \\
v^{l}_{i}
\end{array}
\right),
\label{eq_BdG}
\end{eqnarray}
with $\tilde\Delta_{ij}=\Delta_{ij}+\Delta_{ji}$.
For a given value of the chemical potential $\mu$ one solves 
the system of equations (\ref{eq_BdG})
self-consistently together with (\ref{ord_pars}) for the energies $E^l$ and 
the functions $u^l_i,v^l_i$. For a system on a square lattice
of  $n\times m$ sites the size of the matrix is $(2nm)^2$.  
If the total carrier concentration $n=n_f+2n_b$ is fixed, one finds
$\mu$ from 
\begin{equation}
n={1\over N}\sum_{i\sigma}\langle\hat{c}_{i\sigma}^{\dagger} 
\hat{c}_{i\sigma}\rangle +
{2\over N}\sum_{i}<\hat{b}_{i}^{\dagger} \hat{b}_{i} >
\end{equation} 
and using  equations (\ref{n_b}-\ref{ord_pars}) and (\ref{l-dens}).
 
The solutions enable  calculation of all 
parameters of interest. In particular the local density of states (LDOS) 
$N(E, i)$ at the site $i$ as a function of energy $E$ is given by 
\begin{equation}
N(E,i)=\sum_l\left[|u^l_i|^2\delta(E-E^l)+|v^l_i|^2\delta(E+E^l)\right].
\label{ldos}
\end{equation}
For the purpose of numerical calculations we shall replace
the Dirac $\delta(E)$ distribution by the Lorentzian 
${{1}\over {\pi}} {{\eta} \over {E^2+\eta^2}}$, with $\eta$ 
being a small smearing parameter. 
The local fermion number density 
$n_{f,i}=n_f(\vec R_i)=\sum_{\sigma}\langle c^+_{i\sigma}c_{i\sigma}\rangle$ 
is found to be
\begin{equation}
n_{f,i}=\sum_l\left[|u^l_i|^2f(E^l)+|v^l_i|^2(1-f(E^l))\right],
\label{l-dens}
\end{equation}
where $f(E^l)=1/[\exp(E^l/k_BT)+1]$ is the Fermi-Dirac distribution function.
Parameter $\chi_i$ is given by
\begin{equation}
\chi_{i} = \displaystyle \sum_l \sum_{<j>}  g_{i j}\left[
-u^l_iv^{l*}_j(1-f(E^l))+u^l_jv^{l*}_if(E^l)\right].
\end{equation}

The above equations are general in a sense that they allow for
the study of both s-wave and d-wave superconductors depending on the
choice of coupling $g_{ij}$. For the d-wave symmetry we  define
the local value of the order parameter at the site $i$ as the staggered 
sum \cite{andersen2006} over the neighboring bonds $j$, 
\begin{equation}
\Delta_{i}=(-\tilde\Delta_{i,i+y}+\tilde\Delta_{i,i+x}
-\tilde\Delta_{i,i-y}+\tilde\Delta_{i,i-x})/4.
\label{loc-d}
\end{equation}
 
The correlation function  $C_{f,h}$ 
between various parameters $f,h$ (in fact, also 
depending on the points of our lattice $f_i$ and $h_i$) 
is defined as \cite{nunner2005}
\begin{equation}
C_{f,h} \left( R \right) = \frac{\displaystyle \sum_{i} 
\sum_{j} \left( f_{i} - f_{average} \right)\left( h_{j} - h_{average} 
\right) }{ \displaystyle \sqrt{\sum_{i} \left( f_{i} - f_{average} \right)^2} 
\sqrt{\sum_{i} \left( h_{i} - h_{average} \right)^2} },
\label{cor-fg}
\end{equation}
where $f_{average}$ and $h_{average}$ are their values averaged
over all sites and $R=|i-j|$ is the fixed distance between 
the sites $i$ and $j$ in the above summations.

In the next subsection we first recall the main
features of the homogeneous boson - fermion model (\ref{BF_hamilt0})
and compare the results for a small cluster with those for the  
bulk homogeneous system.

\subsection{Clean system}

We consider here a  real space version of the BF model  and solve the 
corresponding Bogoliubov - de Gennes equations  first for
the homogeneous system of finite size with the periodic boundary 
conditions and compare the results with those obtained for the bulk
system for the same set of parameters. 
 This serves as a check of the quality of 
calculations for a relatively small size of  the system.  

\begin{figure}
\begin{center}
\hspace{-0.25cm}
\includegraphics[width=5.30cm]{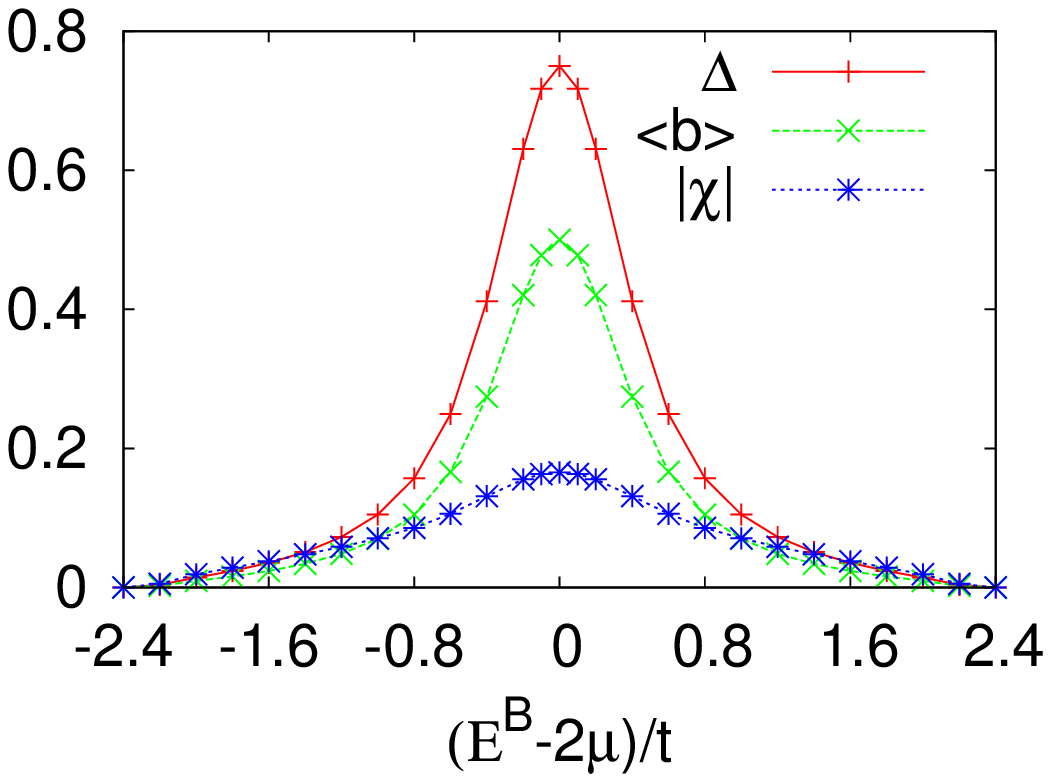}
\hspace{-0.55cm}
\includegraphics[width=5.70cm, height=3.8cm]{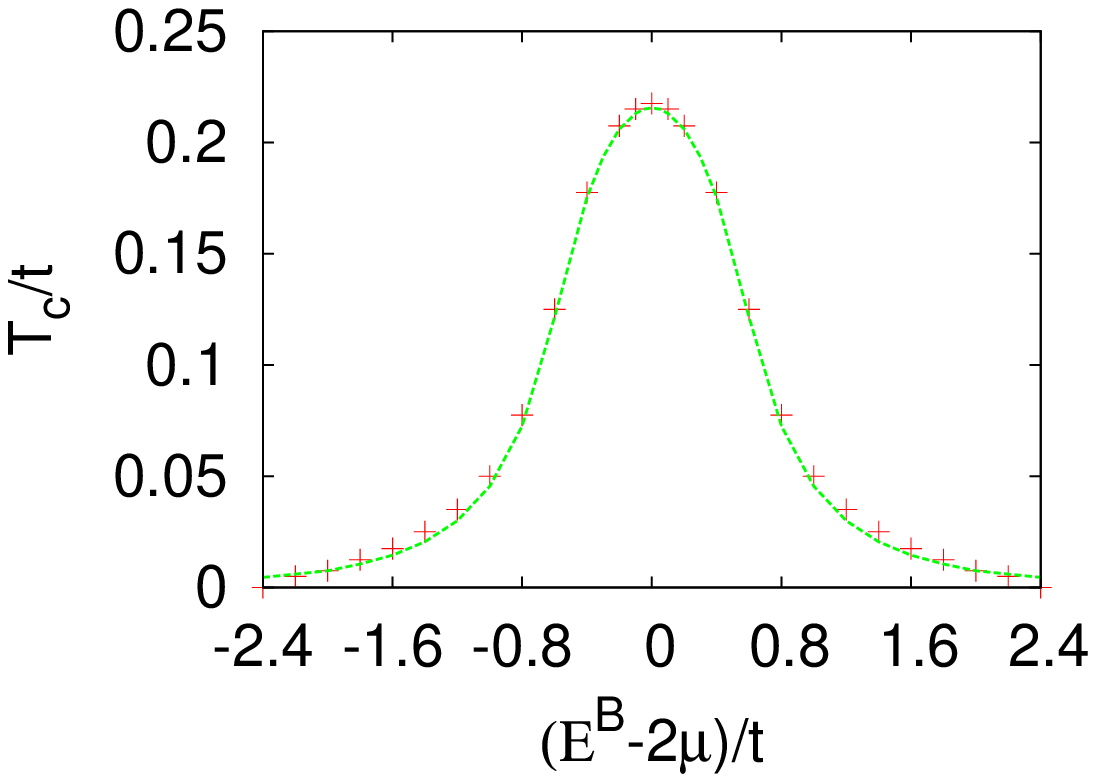}\\
\includegraphics[width=5.30cm]{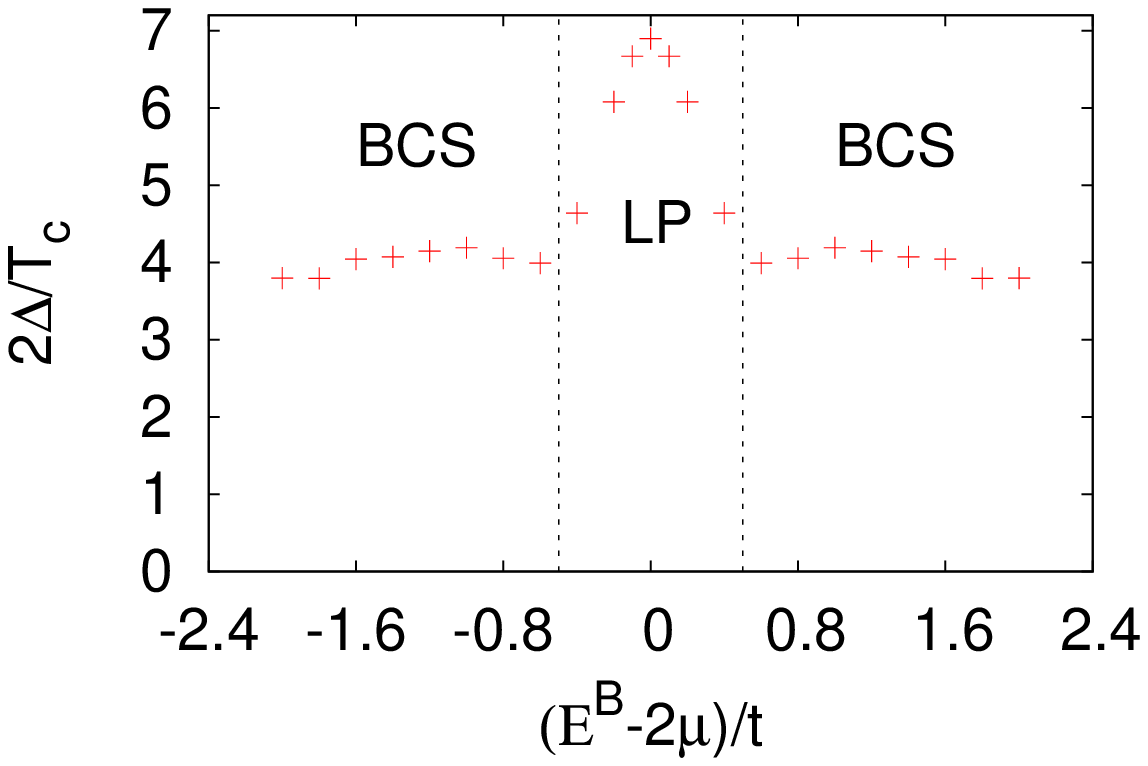}
\includegraphics[width=5.30cm]{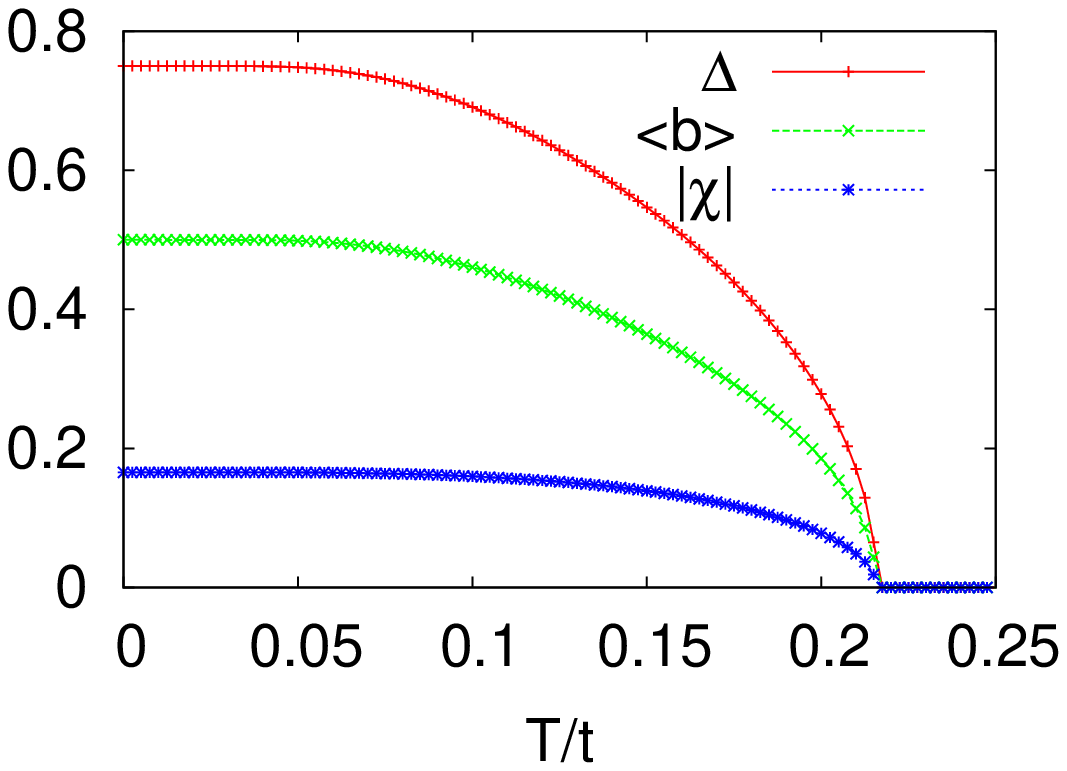}
\caption{(Colour online) The local order parameters $<b_i>$, $|\chi_i|$ and
 ($\Delta_i$) at $T$=$0$  in the units t 
(upper left panel), superconducting transition temperature  ($T_{c}/t$)
(upper right panel) and their ratio  $2 \Delta / T_{c}$ (lower left 
panel) are plotted as the functions  of the boson energy
$(E^B-2\mu)/t$ for a small, homogeneous system of the size $17\times21$ with 
$g$=$0.375t$, $\mu$=$0.0t$, $V^{imp}$=$0.0t$. The lower right panel
shows the dependence of various (coupled) order parameters on temperature
for $E^B$=$2\mu$. The superconducting transition temperature (crosses) 
obtained here
is compared with that calculated for the corresponding bulk system
(dotted curve).
}
\label{fig-clean1}
\end{center}
\end{figure}

Figure \ref{fig-clean1} shows the dependence of the zero temperature 
order parameters $\left< b \right>$, $|\chi|$ and the corresponding
energy gap $\Delta$ (upper-left panel), the critical temperature 
$T_c$ (upper-right panel) and their ratio $2\Delta/T_c$ (bottom left panel) 
on the boson energy level $E^B$ measured with respect to
the chemical potential. In the upper right panel the solid  
curve  represents
the data for the bulk homogeneous system, whereas  
the crosses show the results for the finite cluster.
 These data  obtained for d-wave superconductor by solving 
the Bogoliubov - de Gennes equations 
for the very small system of size  $17\times 21$ sites and  the 
periodic boundary conditions agree very well  
with the results obtained from the mean-field study of the two-
dimensional square lattice in the thermodynamic limit \cite{micnas2004}. 

Small deviations are visible for $E^B\gtrsim t$, when real space 
calculations give slightly larger values.   These are due to the 
finite size effect.
Let us recall that the spectrum of a small system is discrete and  usually 
highly degenerate. The degree of degeneracy, however, is substantially 
reduced if one chooses rectangular shape of the system. This reduces 
number of symmetries in the Bogolubov-de Gennes  equations and lowers degeneracy 
of energies. Best agreement with the  bulk data is obtained for lengths 
$n$ and $m$ being different prime numbers.

We remark that the quantities plotted in Figure \ref{fig-clean1} 
are symmetric with respect to $E^{B}-2\mu=0$. Physically this property
comes from the hard-core nature of the local pairs (LP).  Upon filling 
the bosonic level we  obtain a perfect particle-hole symmetry 
between $n^B$ and $1-n^B$ occupancies. If $\mu=0$  
both these cases equally promote 
the superfluid/superconducting abilities of the coupled boson-fermion 
system (\ref{BF_hamilt}). In particular, let us notice strong 
enhancement of the critical temperature near the half-filled boson 
level which is accompanied by the non-BCS relation $2\Delta/T_c$ 
(marked in lower left panel of Fig. \ref{fig-clean1} as LP)
being a hallmark of the present scenario \cite{ranninger1985}.

The model shows superfluid characteristics which are intermediate between  
those of a BCS regime and those of local bipolaronic pairs 
\cite{micnas1990,alexandrov1994}.
For the homogeneous case, the effective interaction $I_0^{eff}$ 
between fermions
depends on the boson-fermion coupling $g$, the position of the 
bosonic energies $E^B$ with respect to the chemical potential
and hard-core boson concentration $n_B$ 
as  $I_0^{eff}=(g^2/(E^B-2\mu))(1-2n_B)$. This can
be easily seen \cite{micnas2002,micnas2004} by rewriting equations 
(\ref{ord_pars}) -(\ref{eq_BdG}) 
for the clean bulk system in the form of effective BCS equation 
for the fermionic order parameter. 
The coupling $g_{ij}$ induces BCS like pairing  in 
the fermionic subsystem and the Bose-Einstein condensation (BEC) in the
bosonic subsystem. The inherent property of the model is that
superconductivity vanishes  at the same temperature $T_c$ in both  
subsystems \cite{micnas2007}. Temperature dependencies of the  bosonic
 $|<\hat{b}>|$, fermionic  $|\chi_{i}|$ and $\Delta_i$ order
 parameters of the clean system are shown in the lower right 
panel of Figure \ref{fig-clean1}. 

For the vanishing boson-fermion coupling $g_{ij}=0$  model (\ref{BF_hamilt})  
describes the two non-interacting subsystems; dirty fermions and immobile
hard-core bosons in random potential. 
None of the subsystems taken alone undergoes the transition.

\subsection{Remarks on the phase diagram of HTS}

The undoped parent compounds of cuprate HTSs are characterized by 
a non-superconducting ground state, which is antifferromagnetic
insulator. With the increasing doping  
the system starts to be metallic and superconducting.
The superconducting transition temperature T$_c$  increases
with increasing doping,  
attains a maximal value for the optimal doping
and decreases again in the overdoped regime. At the same 
time the normal state properties change dramatically.
The pseudogap state \cite{timusk1999}, which sets in at small doping
disappears and the overdoped system  might behave as  normal 
Fermi liquid. The doping is on the one hand the source of
charge carriers in the model, while on the other hand it introduces
impurities. Judging 
from the appearance of superconductivity for finite 
impurity concentration 
 the plausible assumption would be that impurities 
modify local values of the parameters in the Hamiltonian (\ref{BF_hamilt}) 
in such a way as to promote superconductivity.

We thus assume that oxygen impurities in 
HTS  are the source of carriers (holes) and 
disorder in the system. 
The previous studies of
the disordered boson-fermion model concentrated
on the bulk properties and required the averaging over 
configurations. This was achieved by averaging the
free energy  \cite{pawlowski2001} or using the coherent
potential approximation to calculate the averaged Green
functions \cite{domanski2002,domanski2003b}. 
 The role of disorder is hard to 
overestimate in these materials. It seems that most of their
unusual properties may simply be due to an unprecedented
degree of disorder resulting  from both potential
scattering and random values of other parameters.  
The simultaneous presence of strong correlations in real materials
makes the problem more complicated \cite{wysokinski1999} even at the
mean field level \cite{foot}.

In the two-component model it is the 
boson-fermion scattering which induces superconducting
transition. In this scenario the pseudogap phase is 
 dominated by the  incoherent
 pairs of bosonic character. 
To model  the initial increase  of the superconducting
transition temperature with doping  we assume 
that dopants modulate  bosonic levels $E_i^B$ changing 
them locally from the value well below or above the Fermi level to its
vicinity which allows for efficient fermion-boson
scattering and superconducting instability.  
The proper choice of $E_i^B$ leads
to  positive correlations between the positions of
impurities and the  amplitude of the gap
in the local density of states. 

\section{Results of calculations}

The description of the cuprate superconductors
by the disordered boson-fermion model (\ref{BF_hamilt}) 
allows for a great variety of scenarios. In the
following we shall consider a number of them as
our main goal here is to look at various possibilities
offered by the model at hand.

We    numerically solve the Bogoliubov-de Gennes equations
for small clusters with the size $n\times m$ sites. We use 
periodic boundary conditions, with which the solutions  
 converge to the correct bulk values, even 
for a relatively small system.
 We determine the values of the
order parameter 
$\Delta_{ij}$ for the d-wave superconductivity at each bond $i-j$, 
the  carrier concentration $n_i=n(\vec r_i)$ at each site 
and the local density of states $N(E,i)$. The results are presented 
in the form of maps showing these local parameters. For the d-wave
order parameter we shall use the local ``representation"
of $\Delta_{ij}$  as given by equation (\ref{loc-d}).
In this work we limit the discussion to the analysis of the 
ground state ($T=0K$) properties. In the following we shall 
compare our results with STM data taken at
low temperatures.
Because the energy gap changes very slowly for temperatures T<0.3$T_c$,
as it is visible in Fig. 1 the ground state data
can safely be compared with the experiment

As mentioned before, the  model and approach allow us to study various 
kinds of inhomogeneities. They may result from the local changes
of:

-- $E_i^B$, the position of the boson energy level. As we know from the
study of the model for the clean system, the position of $E_i^B$ with respect
to the chemical potential has a profound effect on the
appearance of superconductivity. 

-- $V_i^{imp}$, which can be interpreted as 
local atomic levels. This term represents the scattering of fermions by
point impurities. 

-- $t_{ij}$, the fluctuating values of hopping parameters. The
 fluctuations of this type are not discussed in this paper.

\subsection{Random bosonic levels supporting 
or suppressing superconductivity}
It is the experimental fact that undoped copper oxides are 
not superconducting. As noted above,    
the existence of coupling between the subsystems $g_{ij}$   
causes superconducting instability. It is also 
evident from Figure \ref{fig-clean1} that for superconductivity to
appear the  bosonic levels $E^B_i$ have to be close 
enough to the Fermi level.
It seems thus natural to expect 
that  doping of parent compounds introduces carriers into
the system and places the boson levels  
in the close vicinity to the Fermi energy. In accordance with such a  
picture we study here a few different (model) scenarios of doping.

First, we consider random $E^B_i$ centers, which we also call
 impurities with parameter values 
supporting superconductivity. It means they move locally   
the boson level towards the Fermi level (c.f. upper
panel of Figure \ref{fig-clean1}). 
Second case is another extreme  in which contrary to
the above expectations impurities move the
boson level out of its initial position in the close
vicinity of $\mu$ and push it up or down
on the energy scale. This makes the boson-fermion scattering 
ineffective due to phase space restrictions. In this scenario
 $E^B_i$ impurities suppress superconductivity. 

The above two scenarios  neglect the effect of 
disorder on the fermion subsystem
{\it i.e.} we put $V_i^{imp}=0$. 
In real systems one expects that the centers with the
boson levels closer to the Fermi energy, at the same time  are
the source of (presumably strong) potential scattering  
for fermions, which in our system
is modelled by $V^{imp}$. Thus as the third scenario 
we consider the system in which the doping will introduce 
randomness in both boson and electron on-site parameters.
In order not to increase a number of parameters we assume \cite{sledz2008} 
that in this case the relation $E^B_i=2V_i$ and $V^{imp}=V_i$
 is valid.

\begin{figure}
\begin{center}
\hspace{0.1cm}
\includegraphics[width=5.50cm,height=4.25cm]{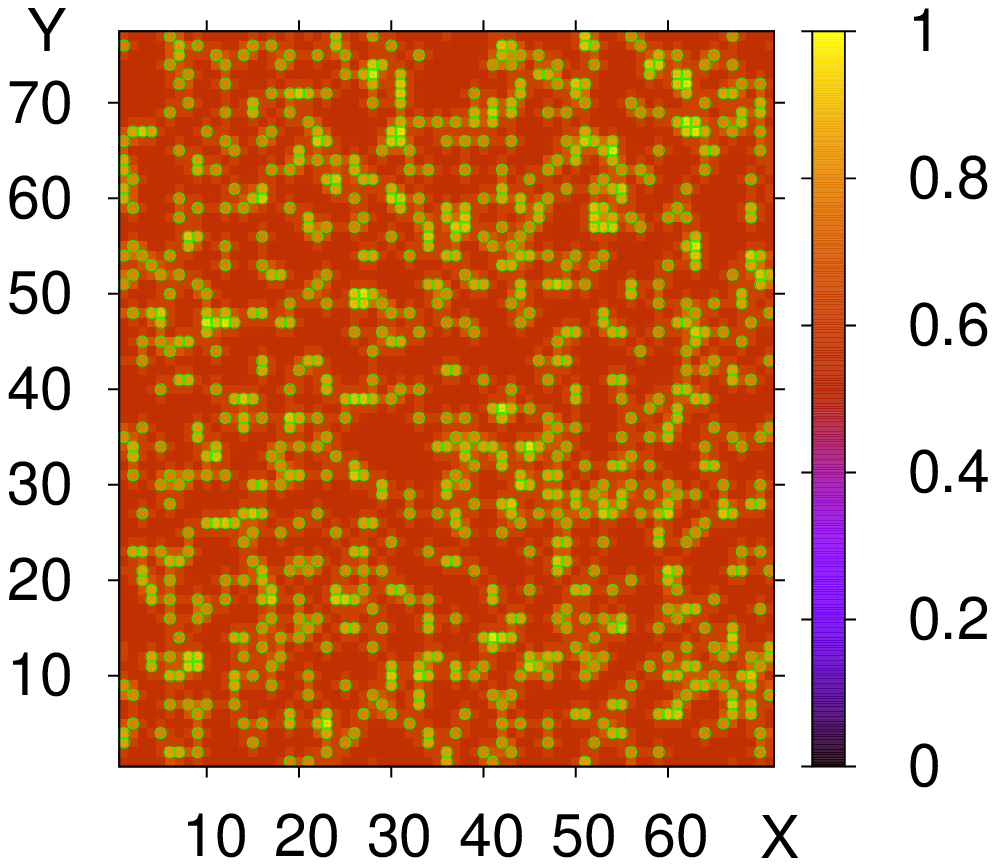}
\hspace{-0.7cm}
\includegraphics[width=5.10cm,height=4.30cm]{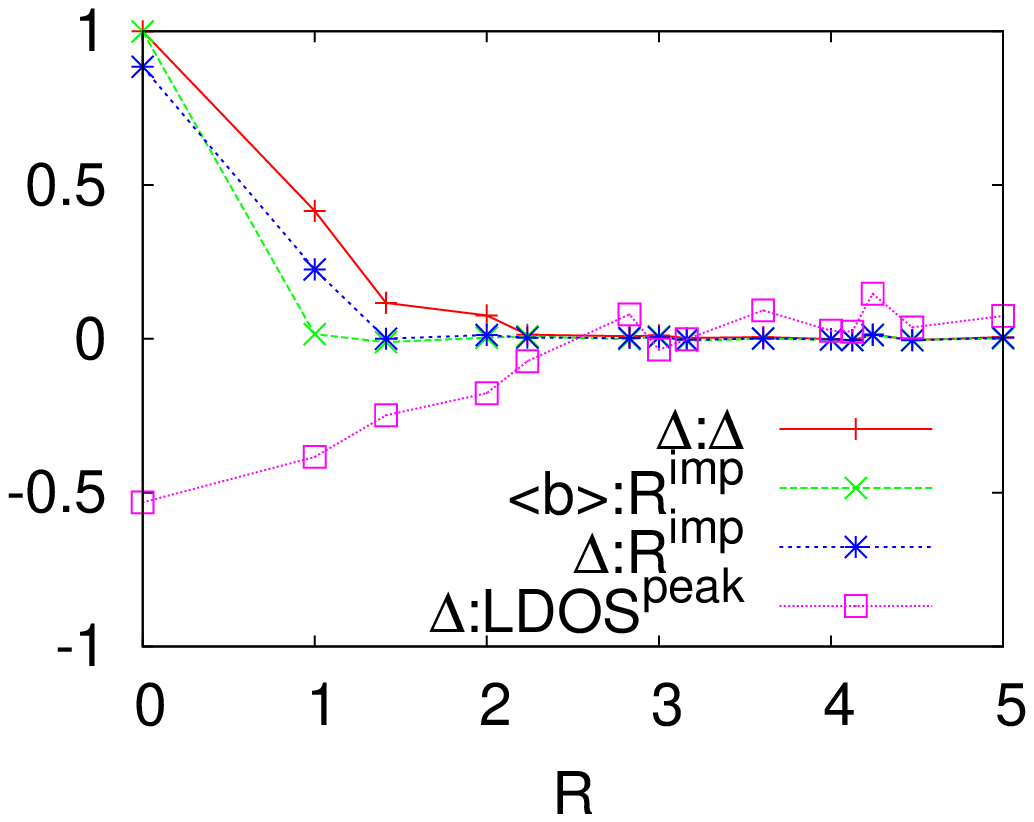}\\
\includegraphics[width=5.0cm]{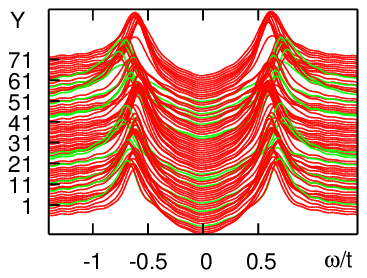}
\includegraphics[width=5.0cm]{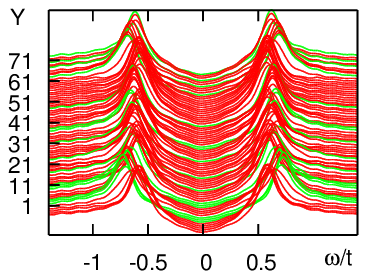}
\caption{(Colour online) Map of the local order parameter $\Delta_i/t$ 
(upper left panel),
the dependence of the correlation functions $C_{f,h}$ (upper right panel),
the STM spectra taken along vertical line at $X$=$53$ (lower left)
and $X$=$67$ (lower right panel). 
The size of the system  is $71\times77$ and there are
$16\%$ of impurities (marked by circles) which change the boson 
levels $E^B_i$ locally from 
the homogeneous value $0.58t$ to $0t$. The other parameters of this 
d-wave superconductor are  $g$=$0.5t$, $V_i^{imp}$=$0$, 
and the total number of carriers $n$=$1.3$.}
\label{fig-Ebi-support}
\end{center}
\end{figure}

\begin{figure}
\begin{center}
\hspace{0.1cm}
\includegraphics[width=5.50cm,height=4.25cm]{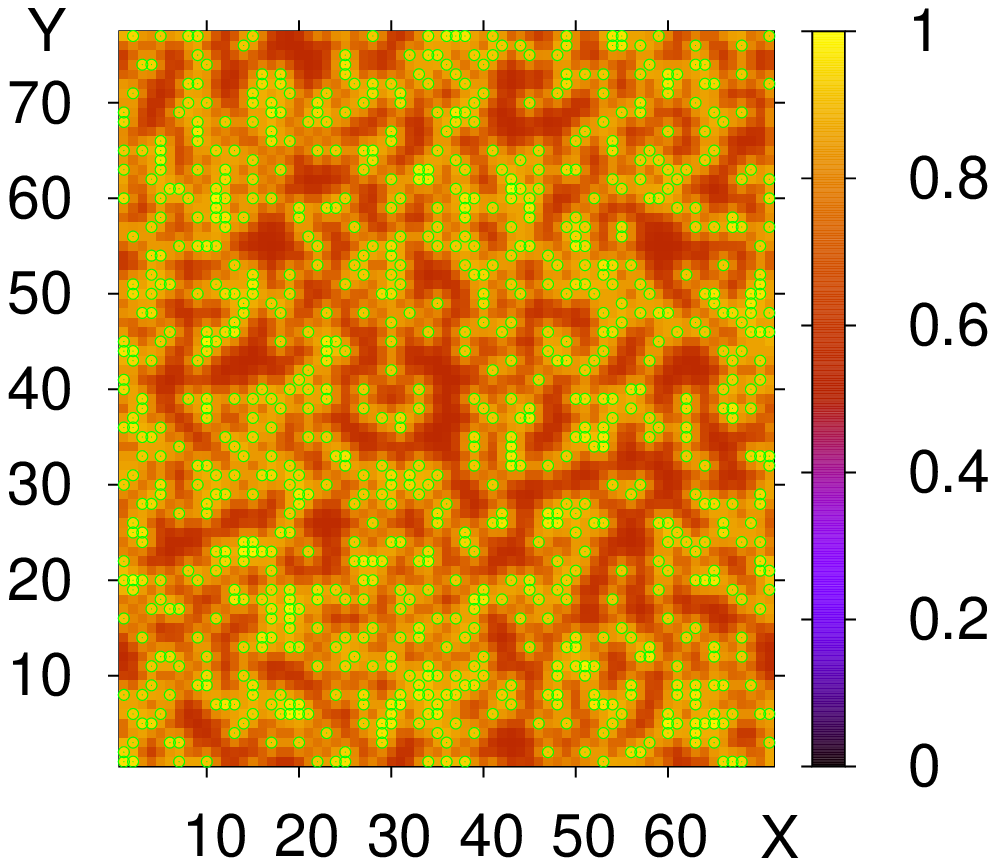}
\hspace{-0.7cm}
\includegraphics[width=5.10cm,height=4.30cm]{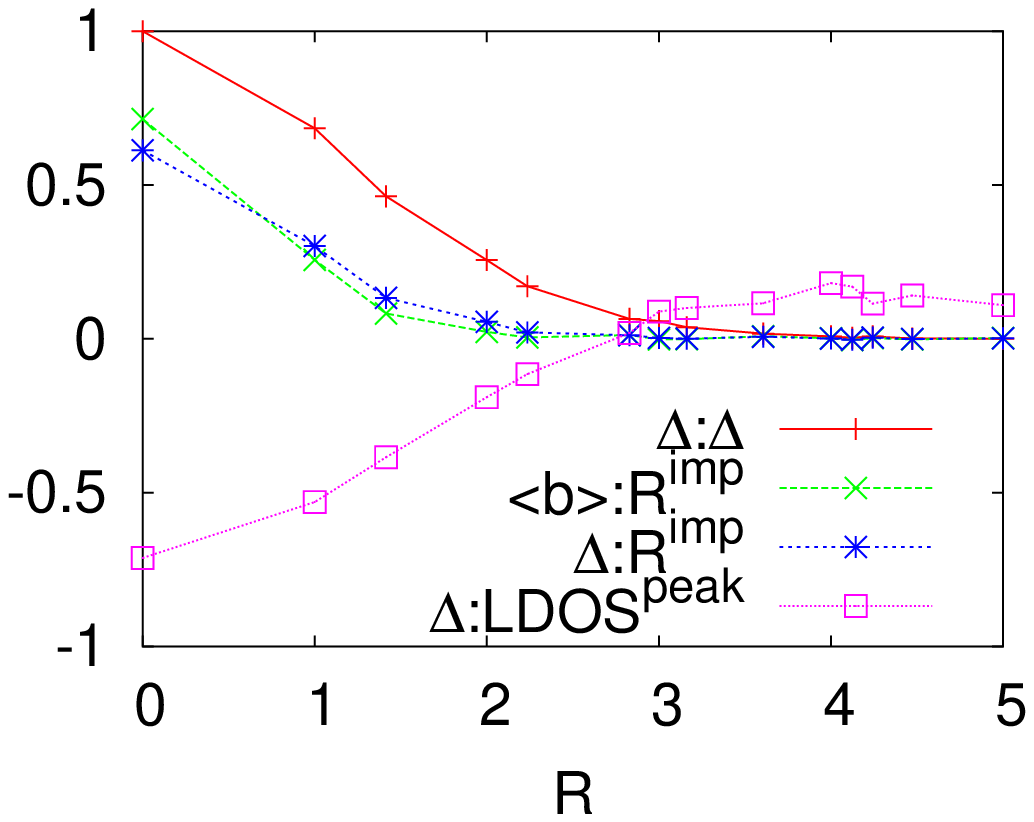}\\
\includegraphics[width=5.0cm]{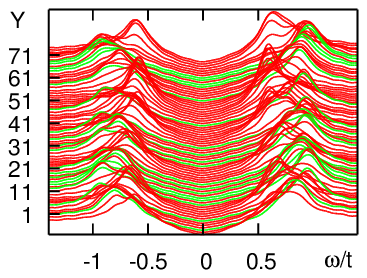}
\includegraphics[width=5.0cm]{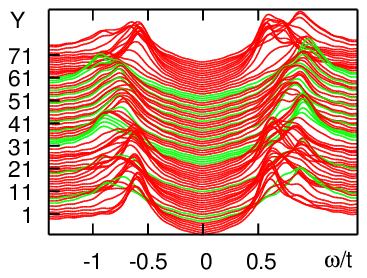}
\caption{(Colour online) The same dependencies as in Figure 
\ref{fig-Ebi-support}
but LDOS is presented for $X$=$17, 43$. We consider here the  
extended impurities which influence the positions of $E^B_i$ levels 
of neighboring sites changing them from the homogeneous value 
$0.58t$ to $0t$ at the impurity site, $0.29t$ at its nearest neighbor site, 
$0.425t$ and $0.5t$ at still further sites.}
\label{fig-Ebi-support-ext}
\end{center}
\end{figure}

\begin{figure}
\begin{center}
\hspace{0.1cm}
\includegraphics[width=5.50cm,height=4.25cm]{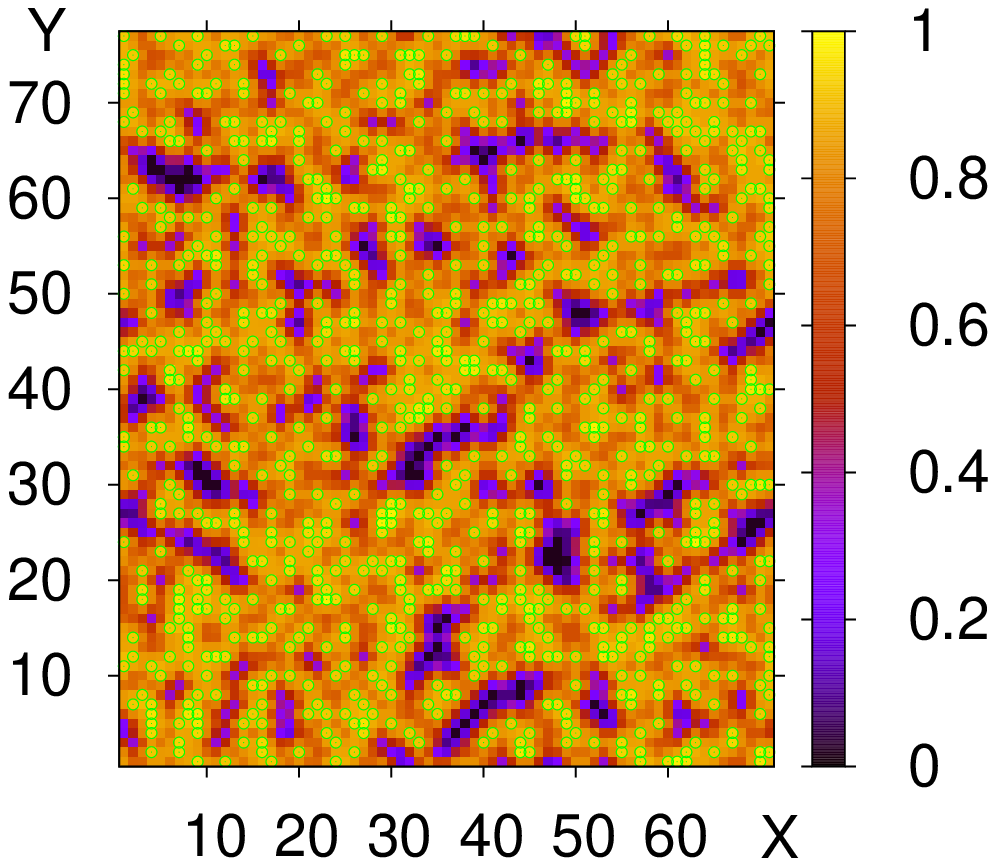}
\hspace{-0.7cm}
\includegraphics[width=5.10cm,height=4.30cm]{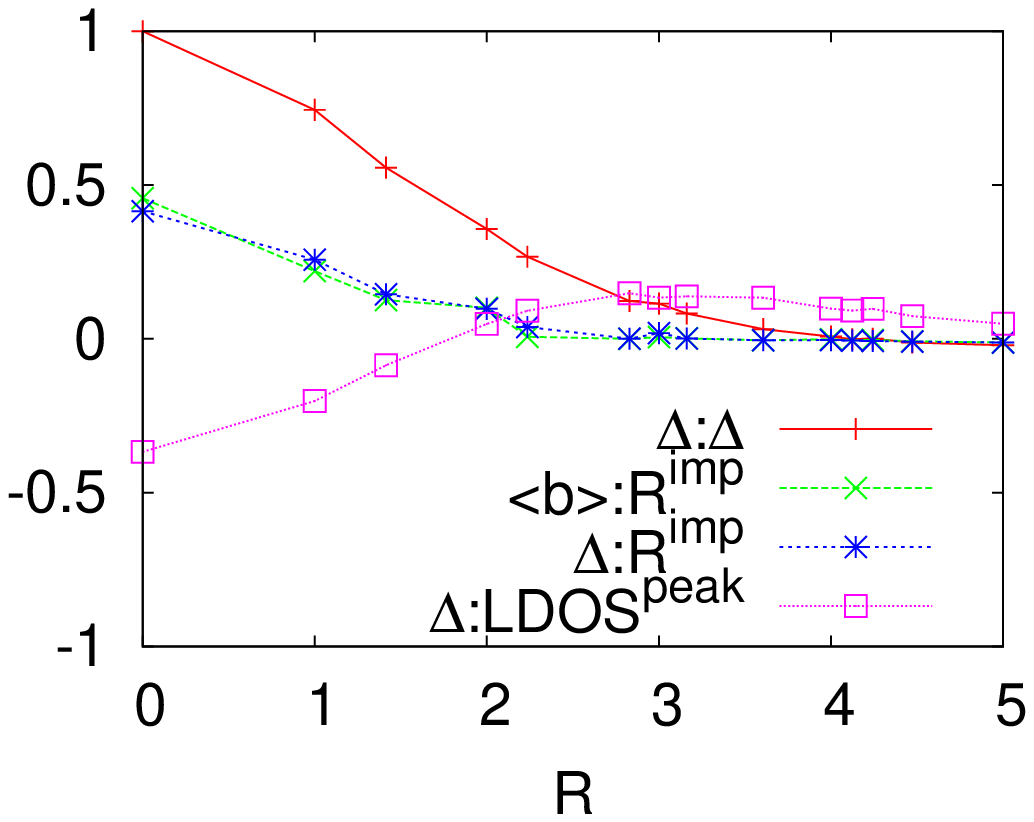}\\
\includegraphics[width=5.0cm]{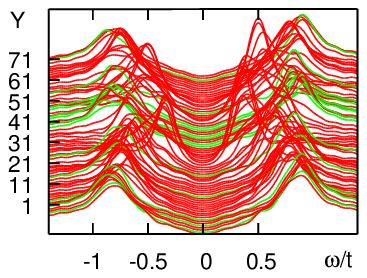}
\includegraphics[width=5.0cm]{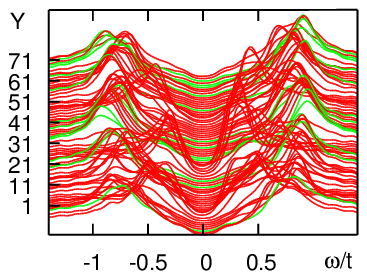}\\
\hspace{0.7cm}
\includegraphics[width=5.50cm,height=4.25cm]{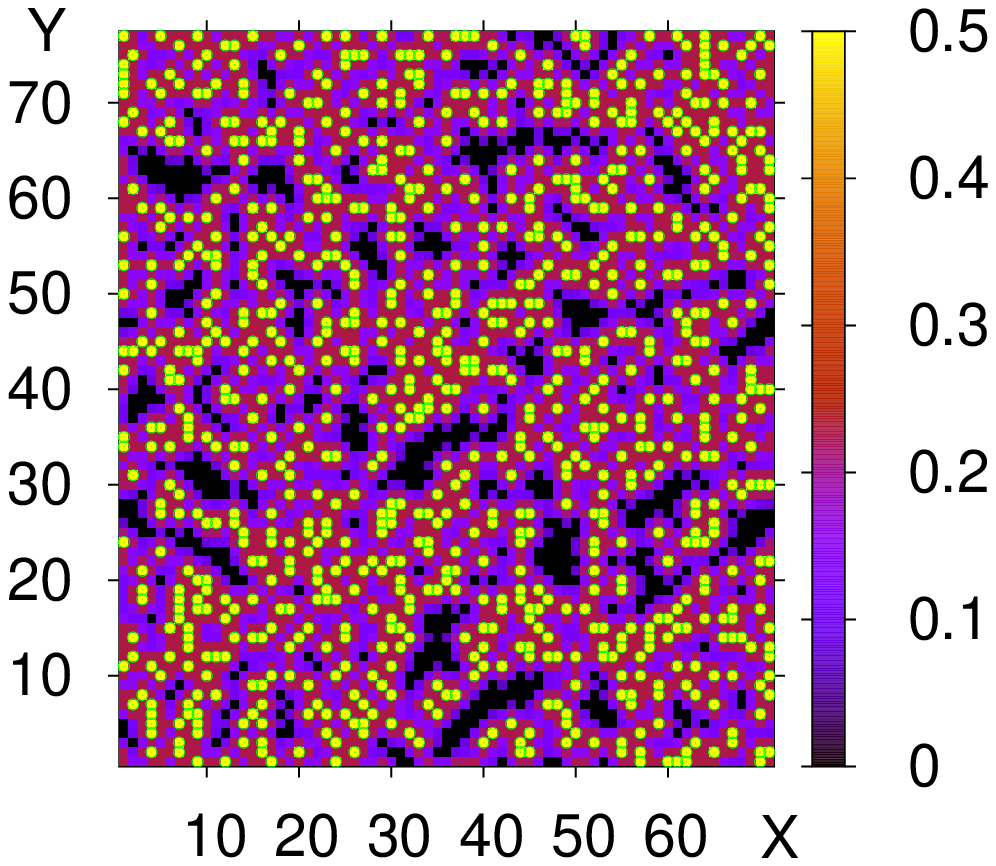}
\hspace{-0.5cm}
\includegraphics[width=5.50cm,height=4.25cm]{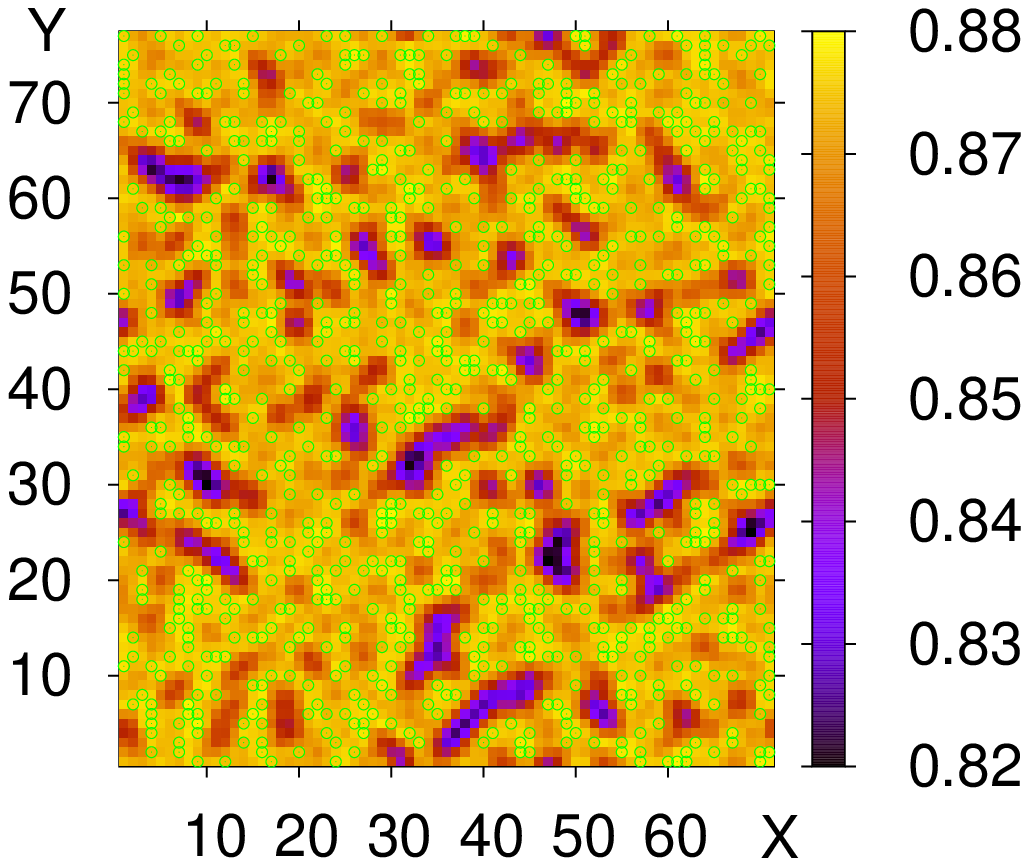}
\caption{(Colour online) Effect of extended impurities on the non-superconducting 
system with $E^B_i$=$4t$. The same dependencies as in 
Figure \ref{fig-Ebi-support-ext} except LDOS are shown for $X$=$26, 33$
and the lower panels show the maps of carrier concentration: bosons 
(left), and fermions (right). The total carrier concentration is kept constant
$n$=$1.3$. The size of the system is $71\times77$ and there are
$16\%$ of impurities (marked by circles on the map of $\Delta_i/t$) 
which change the boson levels $E^B_i$  from 
the homogeneous value $4t$ to $0t$ at the impurity site, $0.29t$ at its 
nearest neighbor site, $0.425t$ and $0.5t$ at still further sites.} 
\label{fig-sup-ext}
\end{center}
\end{figure}

\begin{figure}
\begin{center}
\hspace{0.1cm}
\includegraphics[width=5.50cm,height=4.25cm]{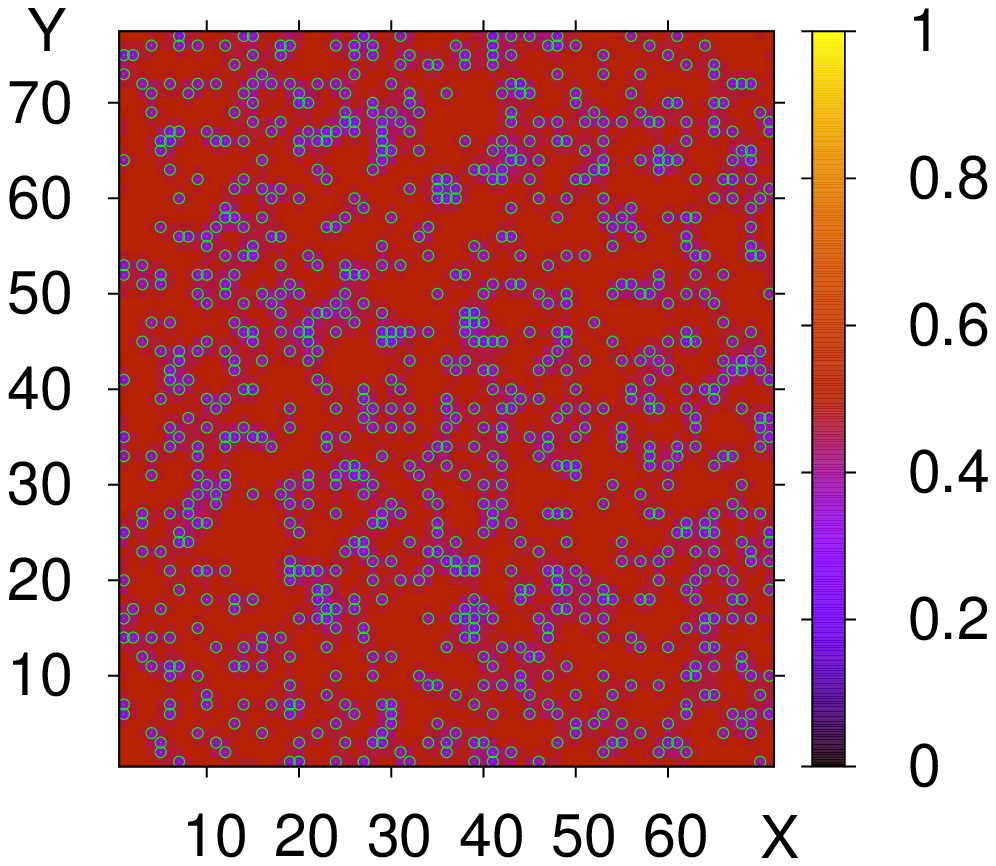}
\hspace{-0.7cm}
\includegraphics[width=5.10cm,height=4.30cm]{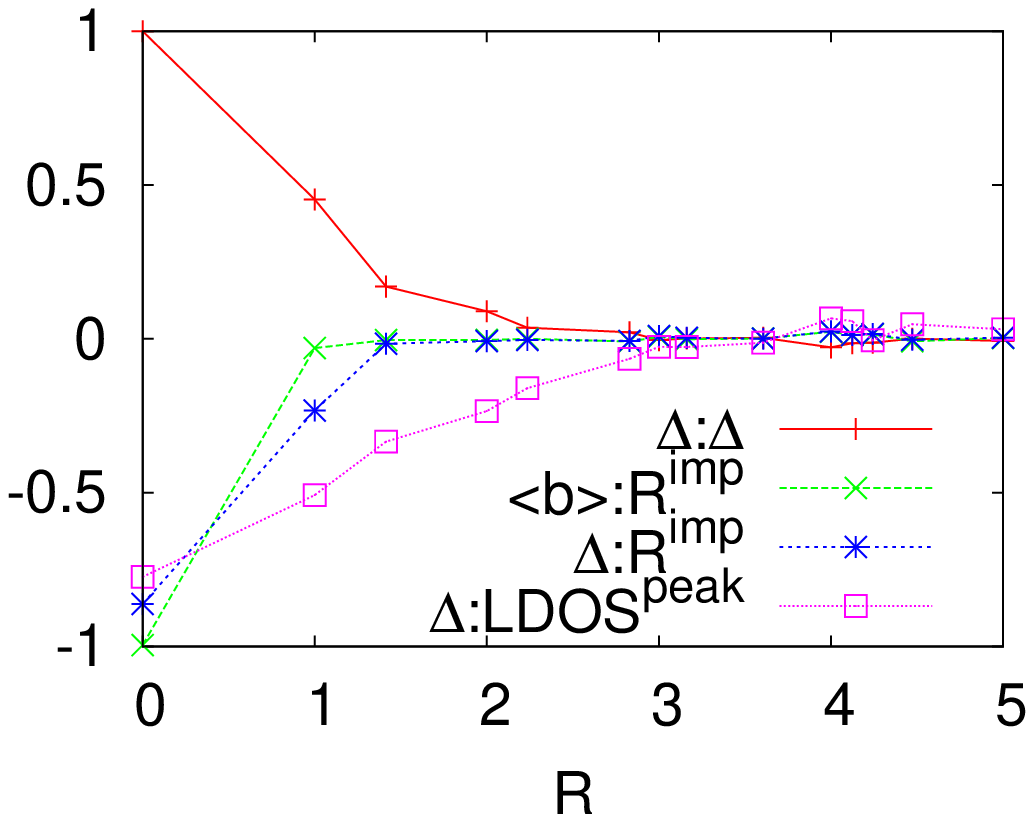}\\
\includegraphics[width=5.0cm]{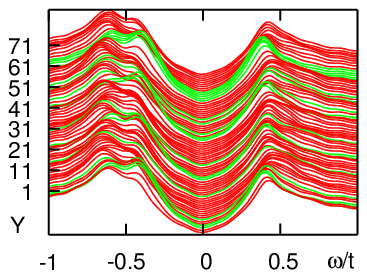}
\includegraphics[width=5.0cm]{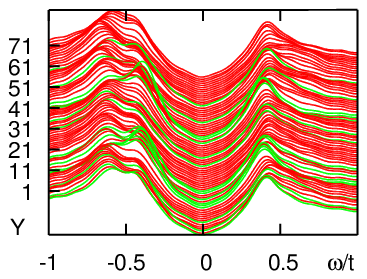}
\caption{(Colour online) Map of the local gap $\Delta_i/t$ (upper left panel),
the dependence of the correlation functions $C_{f,h}$ (upper right panel),
the STM spectra taken along the vertical line at $X$=$29$ (lower left panel)
and $X$=$39$ (lower right panel). 
The size of the system  is $71\times77$ and there are
$16\%$ of impurities (marked by circles on the map of $\Delta_i/t$) 
which change the boson  levels $E^B_i$ locally from 
the homogeneous value $0.58t$ to $1.16t$. The other parameters of the 
d-wave superconductor are   $g$=$0.5t$, $V_i^{imp}$=$0$, 
and the total number of carriers $n$=$1.3$.}
\label{Ebi-suppres}
\end{center}
\end{figure}

It has to be stressed that in principle there exists  
an additional degree of freedom  which 
is related to the spatial extent \cite{andersen2006} at 
which introduction of the impurity 
 into the lattice will modify the parameters of the 
effective model as (\ref{BF_hamilt}). Accordingly we  
discuss short (local) and long range (extended) impurities.
In the latter case the impurity at a given site will
modify the boson energy levels in neighboring and more distant 
sites (see Figs. \ref{fig-Ebi-support}-\ref{fig-Ebi-support-extV}). 
If two extended  impurities happen to influence
the same site their effects simply add up.

In Figure \ref{fig-Ebi-support} we show a
superconducting system with relatively low
transition temperature and introduce local impurities
which move the boson levels $E^B_i$ towards the chemical potential,
thus increasing $T_c$. Next two figures ({\it i.e.} Figure 
\ref{fig-Ebi-support-ext} and \ref{fig-sup-ext}) 
 refer to the extended impurities.
The maps of the local order parameter $\Delta_i$ of the d-wave superconductor 
are shown in the upper left panels. 

The upper right panels of Figures  \ref{fig-Ebi-support-ext} 
and \ref{fig-sup-ext} show 
the distance $R$ dependence of the correlation functions $C_{f,h}$,
as defined in Eq. (\ref{cor-fg}),
for the following parameters ($f,h$): 

(i) local values of the fermionic order parameter 
$\Delta_i$ with positions of the impurities; $R^{imp}$,

(ii) local values of bosonic order parameter 
$\langle b_i\rangle$  with  $R^{imp}$

(iii) $\Delta_i$ with the maximal height of the local density
of states $N(i)$ denoted as LDOS$^{peak}$ in the figures. 
Note, that at each lattice site  the height of $N(i,\omega)$ is the largest at
slightly different energy $\omega$.

(iv) the correlation of $\Delta_i$ with its
value at the distance $i+R$. The extent of this
correlation can be viewed
as a measure of the coherence length $\xi$. For the short
range impurities it is of the order of one lattice spacing
and increases with the range of effective interaction.

The STM spectra shown as the functions of energy (bias)
in  Fig. \ref{fig-Ebi-support} 
are taken along vertical line at $X=53$ (left)
and $X=67$ (right), whereas those in Fig. \ref{fig-Ebi-support-ext}
are for $X=17$ and $X=43$, respectively. 
 The size of the system is $71\times 77$ and there are
16\% of impurities (marked by circles) which change the boson 
levels $E^B_i$ locally from the homogeneous value 
0.58t to 0t (Fig. \ref{fig-Ebi-support}). 
Other parameters of this d-wave superconductor are 
$t_{1}=1.0t$, $g=0.5t$, $V_i^{imp}=0$, 
and the total number of carriers $n=1.30$.

It has to be noted that even without impurities
the homogeneous system with the above parameters 
is superconducting. These additional impurities
increase its $T_c$. 
They also change the values of the local gaps,
which are  visible in the energy dependence of the
local densities of states. 

The correlation   
between the position of the
impurities and amplitudes of the local 
order parameters $\Delta_i$ and $\left<b_i\right>$
in Fig. \ref{fig-Ebi-support} is positive and 
short ranged. It is important to note
slight differences between the two correlations
which are related to strictly local nature of the 
bosonic order parameter  $\left<b_i\right>$ and
a longer range of the fermionic one. 
The summation of $\Delta_{ij}$ over  
the sites $j$ neighboring 
 $i$ to get $\Delta_i$ makes the two local order parameters differ.
 The point-like character of effective interaction is responsible
for the non monotonous dependence of these correlations on the
distance. It is smooth for the more realistic case of 
longer range impurities shown in Figures \ref{fig-Ebi-support-ext}
and  \ref{fig-sup-ext}.

 The correlations
between the local densities of states (at highest values) and
the values of $\Delta_i$  are negative. 
These quantities 
are anti-correlated. This is true for both local, Figure \ref{fig-Ebi-support}, 
and extended Fig. \ref{fig-Ebi-support-ext} impurities. 
The correlation virtually vanishes (Fig. \ref{fig-sup-ext}) 
for the same kind of extended impurities 
introduced into non-superconducting system 
which is here modelled by the position of bosonic levels very distant 
$E^B=4t$  from the Fermi energy 
({\it cf.} upper left panel in figure (\ref{fig-clean1})).

The anti-correlation is easy to understand. Close
examination of the data indicates that 
  each time LDOS is measured for a site where the order parameter is large,
the coherence peaks are relatively broad, whereas they are sharp and 
high in the places with very low or even zero 
value of the order parameter.
For the interpretation of the experiments 
this finding means that each time the STM tip is scanning the 
region of low (or even zero) order parameter it finds sharper edges 
of the spectra  and lower apparent gap 
(defined as the peak to peak distance  in 
the local density of states). On the contrary, the spectra 
calculated for the patches of the sample with 
large values of the order parameters display
less sharp coherence features. 

These sharp coherence-like features  observed
at the sites with low values of ``effective" interactions $I_i^{eff}$ 
resemble those found  in the nonsuperconducting
region placed inside the superconductor \cite{fang2006} 
and induced by the proximity effect. We believe that the same 
general mechanism of proximity is operating here on the local scale
and between regions with  different values of the order parameters.

Another interesting feature of the
STM spectra is their relative homogeneity at low
energies. This qualitatively agrees with the STM 
data  \cite{cren00,pan01,howald2001,lang02,mcelroy05}
 and theoretically a similar behaviour is also observed 
in the  spectra obtained for the two-orbital 
model \cite{ciechan2009} with local interactions (negative Hubbard U).
In this context, we have to remark that as mentioned earlier, 
calculating LDOS we replaced the Dirac delta function by  
the Lorentzian of width $\eta$. In all calculations
reported here we have taken $\eta =0.04t$. This 
necessarily slightly smears
spectra at low energies and cuts the singularities at
the gap edges. 

The comparison of the  correlations $C_{f,h}$ observed in 
Figures \ref{fig-Ebi-support},
\ref{fig-Ebi-support-ext} and \ref{fig-sup-ext}
with those measured experimentally 
\cite{cren00,pan01,howald2001,lang02,mcelroy05} clearly 
indicates the validity of the scenario in which 
doping of the non-superconducting medium introduces bosonic 
impurities in a close vicinity 
of the Fermi level. 

In the above calculations the total concentration 
of carriers is fixed and equals  
$1.3$. However, due to disorder the local values of
boson  $n_b$ and fermion $n_f$ numbers vary from point to point.
The lower panels in Figure \ref{fig-sup-ext} show
colour coded maps of $n_b$ (left panel) and $n_f$ (right panel).
These data as well as similar maps for other cases
with the  fixed carrier concentration show that
fluctuations of bosons are much stronger than those of fermions.
In Figure \ref{fig-sup-ext} the fermion concentration
 varies between 0.82 to 0.88, whereas $n_b$ changes from 0 to 0.5.
The relative fluctuations depend on the kind of impurities
in the system, but as a general rule fermions fluctuate
more weakly than bosons. In most cases studied and in agreement
with the behaviour observed in Figure \ref{fig-sup-ext} the
maps of concentrations show a large degree of correlations 
with the maps of order parameter.  

The results obtained for the second scenario 
according to which we introduce into 
the system bosonic  impurities suppressing superconductivity
are illustrated in Figure \ref{Ebi-suppres}. 
It refers to point-like impurities, which change the
value of the $E^B_i$ level only at 
the impurity site. On the map of local gap 
one notices the diminishing of the
superconducting gap at the impurity sites and
in their close vicinity. The correlations between the
gap value and the position of impurities take on negative
values. LDOS as the function of $\omega$, shown in the lower panels
for two cuts along  X=29 (left) and X=39 (right)
show  smaller fluctuations of the gap than those 
observed for the previous scenario with bosonic 
impurities supporting the superconductivity. The 
appreciable asymmetry of $N(i,\omega)$ is due to the fact that for
the parameters chosen the Fermi level of
the impure systems has moved  to the close vicinity of the Van Hove
singularity. 

This scenario is {\it not} consistent
with experimental findings as it leads to
different than observed correlations between local gap and position
of impurities.

\subsection{Random scatterers}
As mentioned earlier we shall assume here that
impurities  supporting superconductivity by
moving the boson levels closer to the
Fermi level at the same time induce potential
scattering. Obviously this is the most realistic scenario, 
as in any real system with weak screening properties one
expects charged impurities to influence not only single
particle properties but also  local interactions.

\begin{figure}
\begin{center}
\hspace{0.1cm}
\includegraphics[width=5.50cm,height=4.25cm]{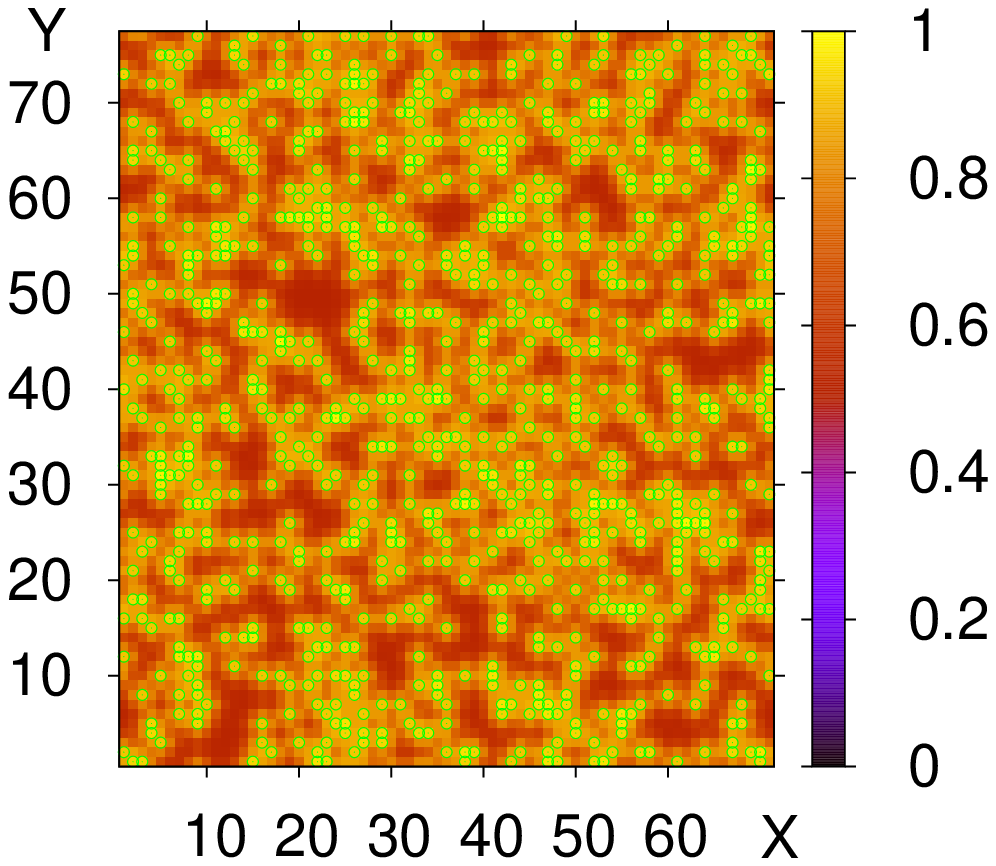}
\hspace{-0.7cm}
\includegraphics[width=5.10cm,height=4.30cm]{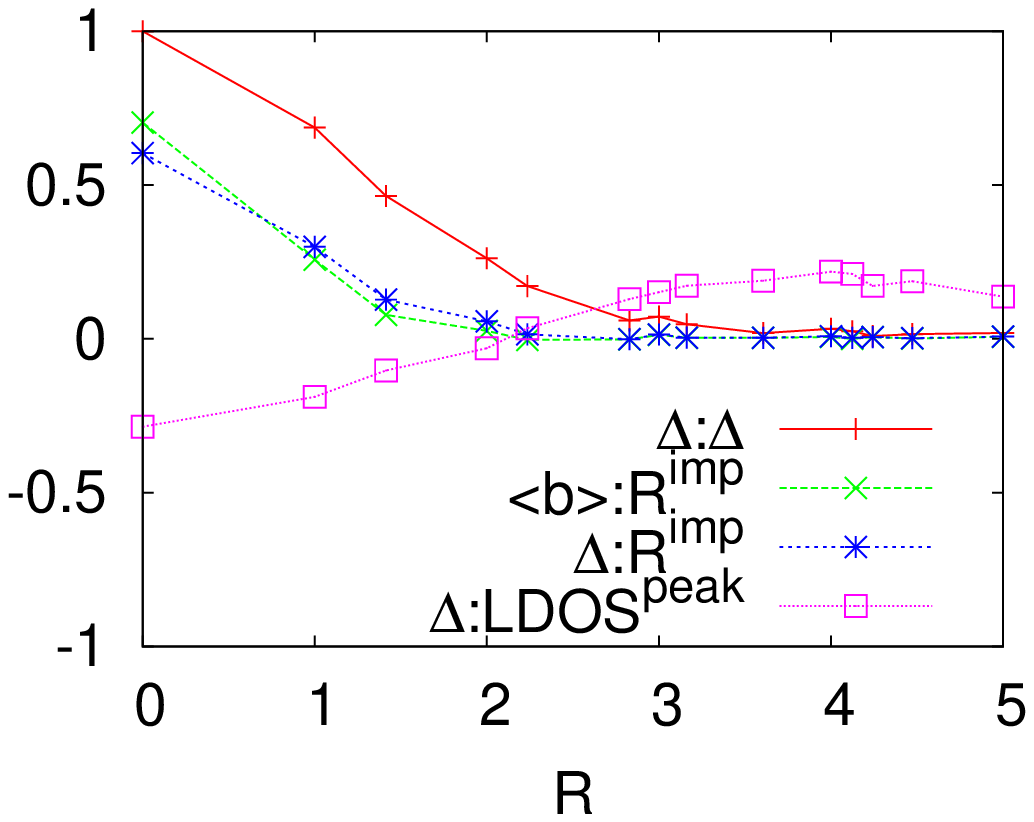}\\
\includegraphics[width=5.0cm]{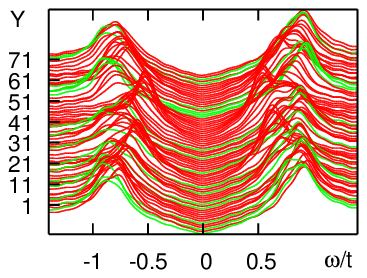}
\includegraphics[width=5.0cm]{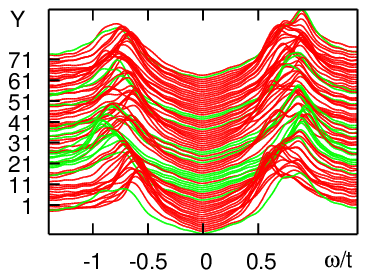}
\caption{(Colour online) Map of the local order parameter $\Delta_i/t$
 (upper left panel),
the dependence of the correlation functions $C_{f,h}$ (upper right panel),
the STM spectra taken along the vertical line at $X$=$23$ (lower left panel)
and $X$=$61$ (lower right panel). The impurities are  
extended and influence the positions of $E^B_i$ levels 
changing them from the homogeneous value $0.58t$ to $0t$
in the impurity position, $0.29t$ at its nearest neighbor site,
$0.425t$ and $0.5t$ at still further sites. In a similar way they also
modify local energies from $V^{imp}_i$=$0$ to $V^{imp}_i$=$-0.29$ 
at the impurity site, $-0.145t$ at its nearest neighbor site, 
$-0.0775t$ and $-0.04t$ at still further sites.}
\label{fig-Ebi-support-extV}
\end{center}
\end{figure}
The results are shown in Figure \ref{fig-Ebi-support-extV} for
a system which, when undoped, is characterized by $E^B=0.58t$ and 
low superconducting transition temperature. The impurities
are extended and modify the bosonic levels at  
the central and neighboring sites  moving them from 
homogeneous value 0.58t to
0t at the impurity position, 0.29t at its nearest neighbor site, 0.425t and
0.5t at still further sites, as in Fig. \ref{fig-Ebi-support-ext}. 
We assume that they also
modify the local ``atomic" energies from $V_i^{imp}=0$ to $V_i^{imp}=-0.29$ 
at the impurity, -0.145t at its nearest neighbor site, -0.0775t and
-0.04t at still further sites. As a result LDOS show strong variations 
from point to point as can be seen in the lower panels 
of Figure \ref{fig-Ebi-support-extV}. 

The correlation 
$C_{f,h}$ between the positions of impurities and the local values
 of the gap is positive and around 0.4 for a small distance. Again it is lower
 than the correlation of the bosonic order parameter due to
the on-site nature of the latter. LDOS shows strong variations 
 from site to  site which are connected with the combined effect
 of changes in the local values of the effective interactions
 and potential scattering. 
It is the potential scattering which suppresses  
sharp coherence  features in the LDOS. The general observation 
that the peaks in LDOS are sharper at the sites with lower 
value of the gap remains still valid.
 
 The map of  local values of  gaps shows characteristic 
 ``two shoulder" distribution with the maximum of smaller gaps centered
 around 0.6t and for larger gaps around 0.9t. This feature, however,
 is not universal and changes with the configuration and the extent of
 impurities.

\section{Summary and conclusions}

We have studied the two-component 
boson-fermion model in real space  paying 
a special attention to the correlations between
the local properties of the system as measured in the scanning 
tunnelling experiments. We have allowed  the site fermion 
energies and bosonic levels  
 to vary in a random fashion. 
Most attention has been paid to variations of the
bosonic levels $E^B_i$. We propose that they are 
mainly affected in the process of doping of 
HTS parent compounds and this is a main source of
inhomogeneity observed in the STM spectra.

The presence of disorder in the bosonic subsystem provides 
natural explanation for a number of observations on 
HTS. In particular, the local values of order parameters 
are positively correlated with the positions of impurities 
($e.g.$ Figure \ref{fig-Ebi-support-ext}). Our calculations for
short ranged bosonic impurities indicate that the 
sharp BCS like features observed in the STM spectra
might not be signatures of the well developed
superconducting gap, but rather present proximity induced
structures, which appear in the regions with the suppressed
superconducting interactions.

In summary, our study shows the ability of the two-component boson-fermion
model to describe the real space STM measurements (dependence of
the  differential conductance on energy, maps of the local gaps and
the correlations between various  parameters) 
with a high degree of accuracy. We limited ourselves  here to
the case of hard core infinitely heavy bosons.  Including kinetic 
energy of bosons, and taking into account 
next nearest-neighbor fermion 
hopping  will make the model more realistic 
and the results will be presented  elsewhere.
The model with mobile bosons has been studied in 
the  homogeneous systems by {\it e.g.} 
\cite{friedberg1989,lindner2010,mamedov2007,squire2009}
and many other authors. This aspect seem to be important for
the correct description of the phase diagram of HTS. 

In this paper we focused on two - dimensional lattice 
 and mean field theory of superconductivity. Accordingly the questions
about the dimensionality and quantum fluctuation 
around mean field solution may arise. 
We consider the surface layer of the superconductor, which 
is coupled to the bulk. It is easy to incorporate
this coupling, as well as hopping to more distant sites 
into our approach, but this would only
slightly change the numerical results but substantially limit the size
of the clusters studied.

The more important issue of quantum fluctuations \cite{emery1995}
can in principle be addressed in a way similar to that
applied recently to the inhomogeneous t-J model \cite{dubi2007}.
In the context of negative U - Hubbard model the inhomogeneities
have been studied \cite{aryanpour2006} by both BdG and Monte Carlo methods.
The conclusion of the authors \cite{aryanpour2006} is that
the results of the mean - field BdG calculations   
agree with Monte Carlo approach which takes into account some of the
fluctuations neglected by the former method. Thus we expect that our
results remain qualitatively correct. 
Influence of the phase fluctuations for the homogeneous 
version of the two-component(boson-fermion) model has 
been addressed by some  of us \cite{micnas2004,micnas2007} 
indicating the suppression of the critical temperature 
due to phase fluctuations. 
The energy gap (or "pseudogap") is  preserved 
up to T*, which in the studied model roughly corresponds 
to $T_c$ obtained from the mean field 
BdG calculations.   

\ack This work has been partially 
supported by the grant no. N N202 1878 33. KIW would like to thank
members of Max Planck Institute f\"ur Physik komplexer Systeme for
hospitality and fruitful discussions. 

\section*{References}
%--------------------------References-----------------------------


\begin{thebibliography}{99}
\bibitem{foot} 
For considered $d$-wave symmetry the on-site Coulomb 
repulsion is not destructive.
\end{thebibliography}
\end{document}